\documentstyle[aps,preprint,epsfig,tighten]{revtex} 

\def\v1{\vspace{1cm}} 
\def\be{\begin{equation}} 
\def\ee{\end{equation}} 
\def\bc{\begin{center}} 
\def\ec{\end{center}} 
\def\vh{\varphi} 
 
\newcommand{\bea}{\begin{eqnarray}} 
\newcommand{\eea}{\end{eqnarray}}

\begin{document} 
\draft 
 
\preprint{ 
\parbox[t]{45mm} 
{MPG-VT-UR 211/00} 
} 
\title{Conformal Cosmology and Supernova Data} 
\author{Danilo Behnke, David Blaschke} 
\address{Department of Physics, University of Rostock, 18051 Rostock, Germany} 
\author{Victor Pervushin, Denis Proskurin} 
\address{Bogolubov Laboratory for Theoretical Physics, Joint Institute for 
Nuclear Research\\ 141980 Dubna, Russia} 
\maketitle 
\begin{abstract} 
We define the cosmological parameters $H_{c,0}$, $\Omega_{m,c}$ and 
$\Omega_{\Lambda ,c}$ within the Conformal Cosmology as obtained by the 
homogeneous approximation to the conformal-invariant generalization 
of Einstein's General Relativity theory.  
We present the definitions of the age of the universe and of the luminosity 
distance in the context of this approach.
A possible explanation of the recent data from distant supernovae Ia without 
a cosmological constant is presented. 
\pacs{PACS number(s):11.25.-w, 04.60.-m, 04.20.Cv, 98.80.Hw} 
\end{abstract} 
\section*{Introduction} 
 
Now there is a very interesting situation in the modern observational 
cosmology stimulated by new data on the distance-redshift 
relation published by the supernova cosmology project (SCP) \cite{snov} and on 
the large-scale structure of the microwave background radiation \cite{10}. 
The SCP data point to an accelerated 
expansion of the universe and have stimulated new developments within 
the standard cosmological model. 
The old naive version of this model with the dust dominance was not sufficient 
to explain these new data. New fits of the modern data in the framework of the 
standard Friedmann-\-Robertson-\-Walker (FRW) model were forced to introduce 
a nonvanishing $\Lambda$-term \cite{9}. The occurence of this term has also 
been interpreted as due to a new form of matter called "quintessence" \cite{9}.
What is the origin of the "quintessence" and why does its density 
approximately coincide with that of matter (luminous plus dark one) at the 
present stage? 
Present theories - containing a scalar field - can describe the data by 
fitting its effective potential \cite{12} but cannot answer these questions. 
 
One of the interesting alternatives to the standard FRW cosmological model is 
the Jordan-Brans-Dicke (JBD) scalar-tensor theory \cite{ps1,13} with two 
homogeneous degrees of freedom, the scalar field and the scale factor. 
Another alternative is the conformal-invariant version of General
Relativity (GR) based on the scalar dilaton field and the geometry of 
similarity (following Weyls ideas \cite{we}) developed in 
\cite{ps1,pr,grg,plb,pct}. 
This dilaton version of GR (considered also as a particular case of the 
Jordan-Brans-Dicke scalar tensor theory of gravitation~\cite{jbd}) 
is the basis of some speculations on the unification of Einstein's gravity 
with the Standard Model of electroweak and strong interactions
~\cite{pr,plb,kl} including modern  theories of supergravity~\cite{kl}. 
In the conformal-invariant Lagrangian of matter, the dilaton scales the 
masses of the elementary particles in order to conserve scale invariance of 
the theory.
However, in the current literature ~\cite{kl} a peculiarity of the 
conformal-invariant version of Einstein's dynamics has been overlooked. 
The conformal-invariant version of Einstein's dynamics 
is not compatible with the absolute standard of 
measurement of lengths and times given by an interval in the 
Riemannian geometry as this interval is not conformal-invariant. 
As it has been shown by Weyl in 1918, conformal-invariant 
theories correspond to the relative standard of measurement of 
a conformal-invariant ratio of two intervals, 
given in the geometry of similarity as a manifold of Riemannian geometries 
connected by conformal transformations~\cite{we}. 
The geometry of similarity is characterized by a measure of changing 
the length of a vector in its parallel transport. In the considered 
dilaton case, it is the gradient of the dilaton $\Phi$~\cite{plb}. 
In the following, we call the scalar conformal-invariant theory 
the conformal general relativity (CGR) to distinguish it from the original 
Weyl theory where the measure of changing the length 
of a vector in its parallel transport is a vector field 
(that leads to the defect of the physical ambiguity of the arrow of time 
pointed out by Einstein in his comment to Weyl's paper~\cite{we}). 
 In the present paper we will apply this approach to a description of  
the Hubble diagram (m(z)-relation) including recent data from the SCP 
\cite{snov} at $z\sim 1$.
We make a prediction for the behaviour at $z > 1$ which drastically deviates 
from that of the standard FRW cosmology with a $\Lambda$ - term. 
We suggest this as a test which could discriminate between alternative 
cosmologies when new data are present in near future. 
 
The present paper is devoted to the definition of the cosmological 
parameters in the Conformal Cosmology by the analogy with the 
standard cosmological model~\cite{pal}.  
To emphasize the mathematical equivalence 
of both cases we try to repeat the standard model definitions
restricting ourselves by the consideration 
of the dust-, curvature-, and $\Lambda$-terms.  
%
%
\section*{Theory and Geometry} 
%
We start from the conformal-invariant theory described 
by the sum of the dilaton action and the matter action 
\be \label{cut} 
W=W_{\rm CGR}+W_{\rm matter}. 
\ee 
The dilaton action is the Penrose-Chernikov-Tagirov one for a scalar 
(dilaton) field with the negative sign 
\be 
\label{wgr} 
W_{\rm CGR}(g|\Phi)= \int d^4x\left[-\sqrt{-g}\frac{\Phi^2}{6} R(g)+ 
\Phi \partial_{\mu}(\sqrt{-g}g^{\mu\nu}\partial_{\nu}\Phi )\right]~. 
\ee 
The conformal-invariant action of the matter fields can be chosen in the form 
\be 
\label{smc} 
W_{\rm matter}=\int d^4x\left[{\cal L}_{(\Phi=0)} 
+\sqrt{-g}(-\Phi F+\Phi^2B-\lambda \Phi^4)\right]~, 
\ee 
where $B$ and $F$ are the mass contributions to the Lagrangians of the vector 
($v$) and spinor ($\psi$) fields, respectively, 
\be \label{67} 
B=v_i (y_v)_{ij}v_j~;~~ 
F=\bar\psi_{\alpha} (y_s)_{\alpha\beta}\psi_{\beta}~, 
\ee 
with $(y_v)_{ij}$, and $(y_s)_{\alpha\beta}$ being the mass matrices of 
vector bosons and fermions coupled to the dilaton field. 
The massless part of the Lagrangian density of 
the considered vector and spinor fields is denoted by ${\cal L}_{(\Phi=0)}$. 
The class of theories of the type ~(\ref{cut}) includes the 
superconformal theories with supergravity~\cite{kl} and the standard model 
with a massless Higgs field \cite{plb} as the mass term would violate the 
conformal symmetry. 
%
\section*{Homogeneous Approximations} 
%
In the Conformal Cosmology, the evolution of a universe is described 
by the scalar dilaton field which can be decomposed into a homogeneous, time 
dependent component and fluctuations, 
$\Phi (T,x)=\vh(T)+\chi(T,x)$. 
In the homogeneous approximation we neglect the fluctuations and start with 
the line element of a homogeneous and 
isotropic universe, which is described by the conformal version of the 
Friedmann-Robertson-Walker (FRW) metric without the scale factor, 
as it disappears due to conformal invariance: 
\begin{eqnarray} 
(ds)^2_c  = g_{00}(t) dt^2 -  \Bigg[ {dr^2 \over 1-k_cr^2/r_0^2} +  
r^2 \Big( d\theta^2 
+ \sin^2\theta \, d\phi^2 \Big) \Bigg]~. 
\label{FRWmetric} 
\end{eqnarray} 
We define 
\be 
dT = \sqrt{g_{00}}dt 
\ee 
 as the conformal time interval. 
From the constraint-type equation $\delta W/\delta g_{00} = 0$ 
we get 
\be\label{phi} 
\vh'^2 = \rho_c\vh + \lambda\vh^4 - \frac{k_c\vh^2}{r_0^2}=\rho_{C}~, 
\ee 
see also \cite{bbp}. 
 
There is a direct correspondence between the conformal cosmology and  
the standard model obtained by the conformal transformations 
\begin{eqnarray} 
dt_f&=&\frac{\vh(T)}{\vh(T_0)} dT = \frac{a(T)}{a(T_0)}dT~,\\ 
\rho_f&=&\frac{\rho_C}{a^4(T)}~, \\
\Lambda&=&\lambda \vh(T_0)^4~, 
\end{eqnarray} 
where $a(T_0)=1$, $\vh(T_0)=M_{\rm Planck}\sqrt{{3}/{(8\pi)}}$. 
The Friedmann time and density are denoted by $t_f$ and $\rho_f$, respectively,
$a(T)$ is the scale factor and $T_0$ the present value of the conformal time.

%
\section*{Determination of the Conformal Cosmological Parameters} 
%
We can define the conformal Hubble-constant 
\be 
H_c = \frac{\vh'}{\vh} = \frac{1}{\vh} \frac{d\vh}{dT}~. 
\ee 
and rewrite~(\ref{phi}) to 
\be\label{phi_1} 
H_c^2(T) = \frac{\rho_c}{\vh(T)} + \lambda\vh(T)^2 - \frac{k_c}{r_0^2}. 
\ee 
Applying it to the present time ($T=T_0$), we can write 
\begin{eqnarray} 
1 = \Omega_{m,c} + \Omega_{\Lambda ,c} + \Omega_{k,c}~, 
\label{sum=1} 
\end{eqnarray} 
%
with the dimensionless parameters 
\bea 
\Omega_{m,c} &\equiv & {\rho_c \over \vh_0 H_{c,0}^2}~,\nonumber\\* 
\Omega_{\Lambda ,c} &\equiv & {\lambda \vh_0^2 \over H_{c,0}^2}~,\nonumber\\* 
\Omega_{k,c} &\equiv & -\; {k\over r_0^2H_{c,0}^2}~, 
\label{Omegas} 
\eea 
where $H_{c,0}$ is  the value of $\vh'/\vh$ at the present time. 
So far, we have discussed only one of the independent equations that 
arises among the set given in Einstein's field equations. In order to 
proceed, we need the other. This is in fact equivalent to the 
statement of conservation of matter, which means that the quantity 
$\rho_c$ is constant in time, and the present-day value of the 
dust matter density is 
\begin{eqnarray} 
\rho_o = \rho_c \vh_0~. 
\end{eqnarray} 

From now on we will use dimensionless variables 
instead of $\vh$ and $T$. We define 
\begin{eqnarray} 
y \equiv \vh/ \vh_0 \,, \qquad \tau_c \equiv H_{c,0}(T-T_0) \,. 
\end{eqnarray} 
Using these variables, we can rewrite Eq.\ (\ref{phi_1}) in the 
following form: 
\begin{eqnarray} 
\left(\frac{dy}{d\tau_c} \right)^2 &=& 
\frac{1}{H^2_{c,0}}\left[ 
{\rho_c y\over \vh_0} + {\lambda y^4 \vh_0^2} - {k_c y^2 \over r_0^2} \right] 
\nonumber\\ 
&=&y^2\left[{1\over y}\Omega_{m,c}+y^2 \Omega_{\Lambda,c}+\Omega_{k,c}\right].
\end{eqnarray} 
Eliminating $\Omega_{k,c}$ now using Eq.\ (\ref{sum=1}), we obtain 
\begin{eqnarray} 
\left( {dy \over d\tau_c} \right)^2 = y^2 \Big[1 + \Big( \frac 1y - 1 \Big) 
\Omega_{m,c} + \Big( y^2-1 \Big) \Omega_{\Lambda ,c} \Big] \,, 
\end{eqnarray} 
or 
\begin{eqnarray} 
d\tau_c = {dy \over y\sqrt{1 + \Big( {1\over y} - 1 \Big) 
\Omega_{m,c} + \Big( y^2-1 \Big) \Omega_{\Lambda ,c}}} \,. 
\label{yeqn} 
\end{eqnarray} 
If there was a big bang, $y$ was zero at the time of the bang, i.e., 
at $T=0$. On the other hand, $y=1$ now, by definition.  Integrating 
Eq.\ (\ref{yeqn}) between these two limits, we obtain 
\begin{eqnarray} 
H_{c,0}T_0 &=& \int_0^1 {dy \over y\sqrt{1 + \Big( {1\over y} - 1 \Big) 
\Omega_{m,c} + \Big( y^2-1 \Big) \Omega_{\Lambda ,c}}} \,. 
\label{yage} 
\end{eqnarray} 
This is the equation which shows that the age of the universe is {\em 
not}\/ independent, but rather is determined by $H_{c,0}$, $\Omega_{m,c}$ and 
$\Omega_{\Lambda ,c}$. 
For the special case of a flat, dust universe without cosmological term 
($\Omega_{m,c}=1$, $\Omega_{\Lambda ,c}=0$), we have $T_0=2~H_{c,0}^{-1}$. 
 
Conventionally, one does not use the dimensionless parameter $y$, but 
rather uses the {\em red-shift parameter}\/ $z$, defined by 
\begin{eqnarray} 
1+z \equiv {\vh_0\over \vh} = {1\over y} \,. 
\label{z} 
\end{eqnarray} 
Using this variable, Eq.\ (\ref{yeqn}) becomes 
\begin{eqnarray} 
d\tau_c =  {dz\over \sqrt{(1+z)^2 (1+\Omega_{m,c} 
z) - z(2+z) \Omega_{\Lambda ,c}}} \,, 
\label{zeqn} 
\end{eqnarray} 
so that 
Eq.\ (\ref{yage}) can be written in the following 
equivalent form: 
\begin{eqnarray} 
H_{c,0}T_0 
&=& \int_0^\infty {dz \over \sqrt{(1+z)^2 (1+\Omega_{m,c} 
z) - z(2+z) \Omega_{\Lambda ,c}}} \,. 
\end{eqnarray} 
Later we will discuss what sort of evolution does this equation 
represent. 
In order to discuss the evolution of the universe, let us not 
integrate Eq.\ (\ref{yeqn}) all the way to the initial singularity, 
but rather to any arbitrary time $T$. This gives 
\begin{eqnarray} 
H_{c,0} (T - T_0) 
= \int_0^{y} 
{dy' \over y'\sqrt{1 + \Big( {1\over y'} - 1 \Big) 
\Omega_{m,c} + \Big( y'^2-1 \Big) \Omega_{\Lambda ,c}}} \,. 
\label{intyeqn} 
\end{eqnarray} 
Equivalently, using the red-shift variable, we can write 
\bea\label{lookback} 
H_{c,0} (T_0 - T) 
&=& \int_0^{z} {dz' \over \sqrt{(1+z')^2 (1+\Omega_{m,c} 
z') - z'(2+z') \Omega_{\Lambda ,c}}} \,. 
\eea 
%
\section*{Distance vs redshift relation} 
\begin{figure}[bth]
\begin{center}
\epsfig{figure=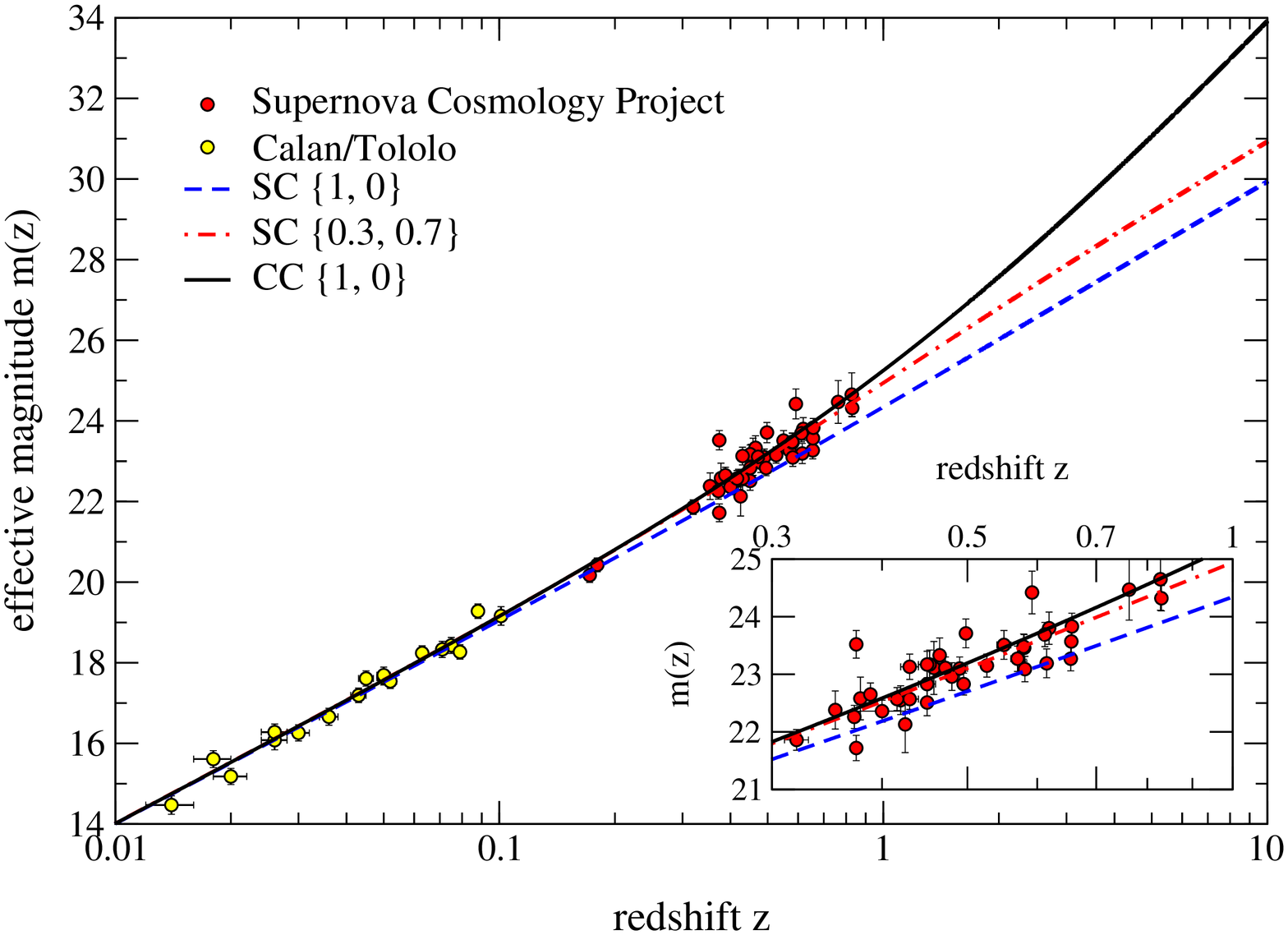,width=12cm,angle=-90} 
\caption{$m(z)$- relation for a flat universe model in Standard and
  Conformal Cosmology. The data points include those
from 42 high-redshift Type Ia supernovae~\protect\cite{snov}.
An optimal fit to these data within the Standard Cosmology requires a 
cosmological constant $\Omega_{\Lambda}=0.7$, whereas in the Conformal 
Cosmology presented here no cosmological constant is needed.
}
\label{fig1}\end{center}\end{figure}

A light ray traces a null geodesic, i.e., a path for which $(ds)_c^2=0$ 
in~(\ref{FRWmetric}). Thus, a light ray coming to us satisfies the 
equation 
\bea 
\frac{dr}{dT} = \sqrt{1-k_c r^2/r_0^2} ~,
\eea 
where $r_0$ is the dimensionless co-ordinate distance introduced in Eq. 
(\ref{FRWmetric}). Using Eqs.\ (\ref{z}) and (\ref{zeqn}), we can 
rewrite it as 
\begin{eqnarray} 
{dr\over \sqrt{1 + \Omega_{k,c} H_{c,0}^2 r^2}} &=& dT\nonumber\\ 
&=& {1\over H_{c,0}} \; {dz\over \sqrt{(1+z)^2 (1+\Omega_{m,c} 
z) - z(2+z) \Omega_{\Lambda ,c}}} \,, 
\end{eqnarray} 
where on the left side, we have replaced $k_c$ by $\Omega_{k,c}$ using the 
definition of Eq.\ (\ref{Omegas}).  Integration of this equation 
determines the co-ordinate distance as a function of $z$: 
\bea
\label{sinn}
&H_{c,0}& r(z) = {1\over \sqrt{|\Omega_{k,c}|}}\times\nonumber\\ 
 &{\rm sinn}& \Bigg[ 
\sqrt{|\Omega_{k,c}|} \int_0^z {dz'\over \sqrt{(1+z')^2 (1+\Omega_{m,c} 
z') - z'(2+z') \Omega_{\Lambda ,c}}} \Bigg]~, 
\eea 
where ${\rm sinn}(x)={\rm sinh}(x)$ for $\Omega_{k,c}>0$, 
${\rm sinn}(x)={\rm sin}(x)$ for $\Omega_{k,c}<0$ and 
${\rm sinn}(x)=x$ for the flat universe with $\Omega_{k,c}=0$. 
The equation~(\ref{sinn}) coincides with the similar relation between the 
coordinate distance and the redshift in Standard Cosmology~\cite{pal}. 
The physical distance to a certain object can be defined in various 
ways. 
For what follows, we will need what is called 
the ``luminosity distance'' $\ell_f$, which is defined in a way that the 
apparent luminosity of any object goes like $1/\ell_f^2$. 
In Standard Cosmology we have 
\begin{eqnarray} 
\ell_f(z) = a_0^2 r(z) / a(z) = (1+z) a_0 r \,. 
\end{eqnarray} 
Thus, 
\bea
&H_0& \ell_f(z) = {1+z\over \sqrt{|\Omega_k|}}\times\nonumber\\
&{\rm sinn}& \Bigg[ 
\sqrt{|\Omega_k|} \int_0^z {dz'\over \sqrt{(1+z')^2 (1+\Omega_m 
z') - z'(2+z') \Omega_\Lambda}} \Bigg] \,. 
\eea 
Any observable distances $\ell_f(z)$ in the Standard Cosmology can be 
converted into observable distances $\ell_c(z)$ in the Conformal Cosmology 
by the conformal transformation 
\be 
d \ell_c = (1+z)d \ell_f(z)~. 
\ee 
Considering the flat universe with $\Omega_\Lambda = 0$ in the dust stage, 
we get in the Standard Cosmology 
\be 
\ell_f(z) = 2 [(1+z) - \sqrt{1+z}]~. 
\ee 
For comparison we obtain for the dust case in the Conformal Cosmology 
\be 
\ell_c(z) = (2z+z^2)-2/3[(1+z)^{3/2}-1]~, 
\ee 
see Fig.~\ref{fig1}. For plotting we used the well known $m(z)-$relation given 
by $m(z) = 5 \log{[H_0\ell(z)]} + {\cal M}$~, where ${\cal M}$ is a constant. 
%
\section*{Conclusion} 
%
We have defined the cosmological parameters $H_{c,0}$, $\Omega_{m,c}$ and 
$\Omega_{\Lambda ,c}$ in the Conformal Cosmology. 
We have shown how the age of the universe depends on 
them. 
The important result of the above derivation of the cosmological
parameters in the Conformal Cosmology is the fact, that we can fit the 
recent Supernova data at $z\sim 1$ in the Hubble diagram (m(z)-relation) in a
simple dust case very well and therefore do 
not need a cosmological constant. Furthermore we gave a prediction for 
the behaviour at $z>1$ which deviates from the standard FRW cosmology
with a non-vanishing $\Lambda$-term. 
Within this model we suggest as a possible explanation of the recent data 
from the Supernova Cosmology Project a static (nonexpanding) flat universe 
where the apparent ``acceleration'' stems from the evolution of the scalar 
dilaton field in Conformal General Relativity, when applied to cosmology.
%
%
 
%
\end{document}